\documentclass[twocolumn,prd,aps,nofootinbib,floats,floatfix,amsmath,amssymb,longbibliography,secnumarabic]{revtex4-1} 
\usepackage[final]{graphicx}
\usepackage{hyperref}
\usepackage{amsmath}
\usepackage{bbm}
\usepackage{amsfonts}
\usepackage{amssymb}
\usepackage{latexsym}
\usepackage{graphicx}
\usepackage[T1]{fontenc}
\usepackage[english,polish]{babel}
\usepackage{multirow}
\usepackage{float}
\usepackage{url}
\usepackage{slashed}
\usepackage{xcolor} 
\usepackage{mhchem}
\usepackage[utf8]{inputenc}
\usepackage[autostyle, english = american]{csquotes}
\usepackage{verbatim}
\usepackage{mathrsfs}
\usepackage{braket}
\usepackage{setspace}
\usepackage{mathtools}
\usepackage{subfigure}

\newcommand{\be}{\begin{equation}}
\newcommand{\ee}{\end{equation}}
\newcommand{\ba}{\begin{array}}
\newcommand{\ea}{\end{array}}
\newcommand{\bea}{\begin{eqnarray}}
\newcommand{\eea}{\end{eqnarray}}

\newcommand{\Tr}{{\rm Tr}}
\newcommand{\nn}{\nonumber}

\addto\captionspolish{}

\begin{document}

\title{Runaway entanglement between a semiclassical background and parametrically-enhanced quantum fluctuations}
\author{Roxane Th\'eriault and George Zahariade}
\affiliation{Instytut Fizyki Teoretycznej, Uniwersytet Jagielloński, Lojasiewicza 11, 30-348 Kraków, Poland}

\begin{abstract}
We study a system of two quadratically-coupled harmonic oscillators, prepared respectively in the ground state and in a coherent state. This toy model can describe for instance the dynamics of one mode of a quantum field evolving in a harmonically-oscillating classical background. In such situations, it is well-known that, depending on the background oscillation amplitude and coupling strength, the system may exhibit parametric resonance. Using the Feynman-Vernon influence functional, we analyze the interplay between parametric resonance and quantum entanglement by perturbatively computing the {\it purity} of the mixed state resulting from tracing over one of the oscillators. We find a runaway behavior of the quantum entanglement between the two oscillators, as well as a rapid breakdown of the validity of perturbation theory, in regions of parametric instability. This provides a quantitative confirmation of various statements in the literature and is highly relevant to questions pertaining to the validity of the semiclassical approximation in coupled quantum-classical systems.

\end{abstract}

\maketitle

\section{Introduction}

Quantum systems evolving in classical backgrounds are ubiquitous in modern physics, from classical textbook examples of non-relativistic quantum particles coupled to external electromagnetic fields (such as in the Stern and Gerlach experiment) to relativistic quantum fields living in curved spacetimes (such as those responsible for the Hawking evaporation of black holes or cosmological pair production) \cite{Gerlach1922-lz,Hawking1975-bk,PhysRevD.35.2955}. As long as the classical background's dynamics are not affected by the presence of the quantum degrees of freedom, the situation is well understood. However, as soon as {\it quantum backreaction} becomes non-negligible i.e. when there is significant transfer of energy, angular momentum, or any other classically-conserved charge between the two systems, the coupled dynamics become significantly more complex. The crux of the problem lies in the simple fact that there is no fully satisfactory way of determining how the coupling to the quantum degree of freedom will modify the equations of motion of the classical one. Of course, one can attempt a fully quantum treatment of the combined system, but this is often unachievable either for fundamental reasons (as in the case of gravitational backgrounds) or for computational ones (because of the sheer complexity of the system of equations). Two closely related questions therefore naturally come to mind: 
\begin{enumerate}
\item[(i)] To what extent and in what situations can we expect quantum backreaction to be negligible and the classical external field approximation to remain valid?
\item[(ii)] When quantum backreaction cannot be neglected, how can one compute it without resorting to the quantization of the full background$+$excitation system? 
\end{enumerate}

The first question does not have an unequivocal answer. The usual heuristic understanding is that as time evolves, the coupling between classical background and any quantum degrees of freedom present in the problem causes them to become quantum entangled. Eventually, the full system may become so entangled that the background ceases to be describable by a well-defined set of classical variables, and a full quantum treatment becomes necessary. (Indeed, quantum correlations are known to not be fully describable in terms of classical ones.) Exactly when this breakdown occurs (or whether it occurs at all) is unclear but this so-called {\it quantum break time} has been the subject of extensive work in recent years \cite{Dvali_2017, Dvali_2018,Dvali_2019,Dvali_2022,Michel_2023}. One thing is clear, however: quantum entanglement between the background and the other degrees of freedom is central to modeling the potential loss of classicality of the background.

The second question is even more complex. There are several prescriptions that are routinely used in the literature to account for quantum backreaction on classical backgrounds (see e.g.~\cite{Braden:2018tky, Hu:1993qa, Vachaspati:2018llo}). Such {\it semiclassical} methods, although imperfect and sometimes arbitrary, aim to track the dynamics of the full system past its quantum break time.  It is however unclear to what extent their predictions differ or what their regime of applicability is. Recent work by one of the authors \cite{van_Dissel_2025} attempted to investigate this question, focusing largely on one of the most widely-used semiclassical approximations, the so-called {\it mean field} approximation (often known simply as {\it the} semiclassical approximation in a gravitational physics context) whereby any occurrence of a quantum operator in the classical equations of motion for the background is replaced by its average expectation value taken in the relevant Heisenberg picture quantum state. In~\cite{van_Dissel_2025}, it was determined that this approximation breaks down faster when the system undergoes parametric resonance and it was conjectured that this was due to a runaway growth of the quantum entanglement between the background and the quantum degree of freedom. This would in fact be consistent with various statements in the literature claiming that the associated entanglement entropy should grow linearly in time with a rate given by the positive Lyapunov exponents of the system~\cite{Bianchi_2018,Zurek:1994wd}.

In the present work, we aim to put this conjecture on a more solid basis. To do that, we will study the simplest possible model that can be partitioned into two interacting subsystems (which could therefore potentially be entangled) and that exhibits phenomena that can be related to parametric resonance in some classical limit. We shall therefore consider essentially the same model as in~\cite{van_Dissel_2025} describing two quantum harmonic oscillators coupled biquadratically. The corresponding Lagrangian is\footnote{Here and henceforth round brackets are used for parametric dependence while square brackets are used for functional dependence. The quantity $L_0(\dots;\dots]$ depends on mass and angular frequency parameters, but also on the coordinate of the system in configuration space, a function of time itself, hence the mixed notation.}
\bea
\label{lagrangian}
L[x,y]&=& L_0(M,\Omega;x]+L_0(m,\omega;y] +L_{\rm int}[x,y]\nn\\
&\equiv&\frac{1}{2}M\dot{x}^2-\frac{1}{2}M\Omega^2x^2\nn\\
&&\quad+\frac{1}{2}m\dot{y}^2-\frac{1}{2}m\omega^2y^2-\frac{\lambda}{2}x^2y^2.
\eea
The reason why this system is interesting for the task at hand is that, at the classical level, linear perturbations $\delta y$ around the background solution $x(t)=X_0\cos \Omega t$, $y(t)=0$, obey a Mathieu equation
\be
\ddot{\delta y} + \left(\omega^2+\frac{\lambda}{m}X_0^2\cos^2\Omega t\right) \delta y=0,
\label{eq:classicalMathieu}
\ee
which is known to exhibit parametric resonance depending on the exact numerical values of the parameters $\omega$, $\Omega$, and $\lambda/m$ \cite{landau1976mechanics}. This is thus a good testbed to investigate the effects of parametric resonance at the quantum level.

As in~\cite{van_Dissel_2025}, we are thus interested in the natural quantum extension of the above purely classical situation where the dynamics of $x$---the ``classical'' background variable---is approximately harmonic with amplitude $X_0$ (and angular frequency $\Omega$), while $y$---the ``quantum fluctuation'' variable---is assumed to be in its ground state. In order to describe this setup in a fully quantum mechanical way, we initialize the $(x,y)$ system in the product state $\ket{\alpha,0}\equiv\ket{\alpha}\otimes\ket{0}$ of a coherent state $\ket{\alpha}$ for $x$ and a properly defined adiabatic ground state $\ket{0}$ for $y$, then let the system evolve unitarily. 

The parameter $\alpha=X_0\sqrt{M\Omega/2\hbar}$ is defined so that the wavefunction of the coherent state $\ket{\alpha}$ is that of the unperturbed ground state for $x$ displaced to position $X_0$.  In the Schr\"odinger picture and in the absence of interactions such a state evolves rigidly into another coherent state with parameter $\alpha e^{-i\Omega t}$, maintaining the Gaussianity of its wavefunction as well as its width, while the average position $\langle \hat{x}(t)\rangle = X_0 \cos \Omega t$ follows the classical trajectory. It is the closest quantum equivalent to a classical ``ball'' rolling in the harmonic well  \cite{Glauber:1963tx,PhysRevLett.10.277,KlauderSkagerstam}. Of course, this behavior is expected to change when interactions are turned on. As far as the adiabatic ground state $\ket{0}$ is concerned, let us mention here that it will be defined with respect to a {\it renormalised} angular frequency $\omega_r$ depending explicitly on $\lambda$ and $\hbar$, and not simply with respect to the {\it bare} $\omega$. This will be discussed in detail around Eq.~\eqref{eq:renfreq}.  

As mentioned above, time evolution will generically result in the state at time $t$ no longer being separable, which leads to quantum entanglement. The aim is to compute the reduced density matrix of $y$ (obtained by tracing out the $x$ degree of freedom) and use it to find the {\it purity} (and more generally the R\'enyi entropy) of the $y$ subsystem. This provides a satisfactory measure of the amount of quantum entanglement between the $x$ and $y$ subsystems~\cite{Schneeloch:2020zkv}. We shall in particular be using many standard tools from the theory of open (dissipative) quantum systems \cite{Kamenev2023FieldTheory}.

In Sec.~\ref{sec:IF}, we begin by perturbatively calculating the so-called {\it influence functional} \cite{FEYNMAN1963118} of system $x$ which encodes, at the quantum level, the average effect that $x$ exerts on $y$. In Sec.~\ref{sec:eff}, we build upon this result by computing the reduced density matrix for $y$ and exploring the effects of parametric resonance on the magnitude of its quantum fluctuations---more precisely, on the amount of {\it squeezing} system $y$ experiences. In Sec.~\ref{sec:ent}, we proceed by computing the purity of the quantum state of $y$ (as a measure of entanglement) as a function of time. In Sec.~\ref{sec:disc}, we discuss our results showcasing the profound relationship between parametric resonance at the classical level, quantum entanglement, and squeezing. We finally conclude in Sec.~\ref{sec:conclusion}.

\section{Influence functional}
\label{sec:IF}

The Feynman-Vernon influence functional formalism is frequently used in the study of open quantum systems, as it allows one to express the behavior of a system of interest, the {\it test system}, coupled to any number of other quantum systems, generically called {\it interaction systems}, in terms of the variables of that system only \cite{FEYNMAN1963118}. The effect of the interaction systems is then encapsulated in a single functional quantity, the {\it influence functional} (IF), and we shall introduce it below. 

In our case, the test system is $y$ and, as mentioned in the introduction, at $t=0$ we initialize it in state $\ket{0}$ with wavefunction $\psi_0(y)$ (to be fully specified later). Similarly, the interaction system, $x$, is initialized in state $\ket{\alpha}$ with wavefunction $\chi_\alpha(x)$. The object that describes the time evolution of this two-degree-of-freedom system is the {\it density matrix} $\hat{\rho}$ and its matrix elements in the position eigenstate basis at time $T$ are
\bea
\bra{x_1,y_1}\hat{\rho}(T)\ket{x_2,y_2}&=&\bra{x_1,y_1}e^{-\frac{i}{\hbar}\hat{H}T}\ket{\alpha,0}\nn\\&&\quad\times\bra{\alpha,0}e^{+\frac{i}{\hbar}\hat{H}T}\ket{x_2,y_2},
\eea
where $\ket{x,y}\equiv\ket{x}\otimes\ket{y}$ and the unitary time evolution operator is obtained by exponentiating the Hamiltonian operator $\hat{H}$ derived from~\eqref{lagrangian}. This can be rewritten as a path integral
\bea
\label{rhoelem}
\bra{x_1,y_1}\hat{\rho}(T)\ket{x_2,y_2}\!&=&\! \int {\cal D}x{\cal D}x'{\cal D}y{\cal D}y'e^{\frac{i}{\hbar}(S[x,y]-S[x',y'])}  \nn\\
&&\hspace{-3cm} \times\chi_\alpha(x_0)\chi_\alpha^*(x_0')\psi_0(y_0)\psi_0^*(y_0')dx_0 dx_0' dy_0 dy_0', 
\eea
where $S[x,y]\equiv\int_0^T L[x,y]dt$ and the boundary conditions in the path integral read $x(0)=x_0$, $x(T)=x_1$, $y(0)=y_0$, $y(T)=y_1$, $x'(0)=x'_0$, $x'(T)=x_2$, $y'(0)=y'_0$, $y'(T)=y_2$.

Since we will ultimately be interested in calculating different measures of entanglement between the two subsystems $x$ and $y$, we will trace out the $x$ subsystem to obtain the {\it reduced density matrix} $\hat{\rho}_R$, which comes down to setting $x_1=x_2$ and integrating over it in Eq.~\eqref{rhoelem}. We thus obtain its matrix elements in the position eigenstate basis,
\bea
\bra{y_1}\hat{\rho}_R(T)\ket{y_2}&=&\int dx_1 \bra{x_1,y_1}\hat{\rho}(T)\ket{x_1,y_2}\nn\\
&&\hspace{-1cm}= \int \mathcal{D}y \mathcal{D}y'F[y,y']\,e^{\frac{i}{\hbar}\left(S_{0}(m,\omega;y]-S_{0}(m,\omega;y']\right)} \nn \\ &&\times \psi_0(y_0) \psi_0^*(y_0') dy_{0} dy'_{0},
\label{eq:reduced}
\eea
where $S_0(m,\omega;y]\equiv\int_0^T L_0(m,\omega;y]dt$  and the boundary conditions on the $y$, $y'$ path integrals remain unchanged. The quantity $F[y,y']$ is the influence functional and is given by
\bea
\label{eq:IF}
    F[y,y']&=&\int \mathcal{D}x \mathcal{D}x' e^{\frac{i}{\hbar}\left(S_{0}(M,\Omega;x]- S_{0}(M,\Omega;x']\right)}\nn\\ 
&&\hspace{-1.5cm}\times e^{\frac{i}{\hbar}\left(S_{\rm int}[x,y]-S_{\rm int}[x',y']\right)}\chi_\alpha(x_{0})\chi^*_\alpha(x'_{0}) dx_{0} dx'_{0} dx_{1},
\eea
where $S_{\rm int}[x,y]\equiv \int_0^T L_{\rm int}[x,y]dt$, and the boundary conditions on the $x$, $x'$ path integrals are $x(0)=x_0$, $x'(0)=x'_0$, $x(T)=x'(T)=x_1$. At this point it is already clear that the IF, which depends explicitly on the coupling between $x$ and $y$, will encode the way in which the interaction with $x$ impacts the dynamics of $y$ (in some average sense, since the trace operation is equivalent to a sum over final states $\ket{x_1}$ in the scattering process whose amplitude is given by $\bra{x_1,y_1}e^{-i\hat{H}T/\hbar}\ket{\alpha,0}$). The aim of this section is to explicitly compute $F[y,y']$. 

Unfortunately, in spite of the fact that each integral appearing in~\eqref{eq:IF} (be it numerical or functional) is Gaussian, there is no way to compute the IF in closed form since $y(t)$ and $y'(t)$ are in principle generic. We thus proceed perturbatively in the coupling constant $\lambda$. Of course, $\lambda$ is a dimensionful quantity so we have to be more precise, but we will postpone this discussion until Sec.~\ref{sec:ent} and, for the time being, we will simply assume that $\lambda$ is small enough so that the truncation of $F[y,y']$ to second order in $\lambda$ provides a reasonably good approximation of the exact result for some time (at least for the purpose of quantifying the amount of entanglement between the two subsystems). The method used here is not the same as the one used  in~\cite{FEYNMAN1963118} but uses the canonical formalism whenever possible to simplify the computations. We thus Taylor expand terms of the form $e^{\pm \frac{i}{\hbar}S_{\rm int}[x,y]}$ in~\eqref{eq:IF} to second order in $\lambda$ to obtain
\bea
\label{eq:IFtaylor}
F[y,y']&=&\int {\cal D}x{\cal D}x'  e^{\frac{i}{\hbar}\left(S_{0}(M,\Omega;x]-S_{0}(M,\Omega;x']\right)} \nn \\ 
&&\hspace{-1cm} \times \left[ 1 - \frac{i\lambda}{2\hbar} \int_{0}^T (x_t^2 y_t^2-x'^2_t y'^2_t)dt\right. \nn \\ 
&&\hspace{-1cm} \left.- \frac{\lambda^2}{4\hbar^2}  \int_{0}^T \int_{0}^t (x^2_t y^2_t-x'^2_t y'^2_t)(x^2_s y^2_s-x'^2_s y'^2_s) dt ds \right] \nn \\ 
&&\hspace{-1cm} \times \chi_\alpha(x_{0})\chi^*_\alpha(x'_{0}) dx_{0}dx'_0 dx_{1},
\eea
where we have used the shorthand notations $x_t\equiv x(t)$, $y_t\equiv y(t)$ and so forth. The boundary conditions on the $x$ and $x'$ path integrals are unchanged.

To make progress in the computation we introduce the propagation kernel
\be
K_f(x_b,t_b;x_a,t_a)\equiv
\int {\cal D}x\, e^{\frac{i}{\hbar}\int_{t_a}^{t_b}dt\left(\frac{1}{2}M\dot{x}^2-\frac{1}{2}M\Omega^2 x^2+fx\right)},
\ee
where the boundary conditions are $x(t_a)=x_a$, $x(t_b)=x_b$, and $f(t)$ is a forcing term. It is well-known \cite{Feynman:100771} that this can be evaluated in closed form, but we will not need the exact expression.
We may now rewrite~\eqref{eq:IFtaylor} as
\bea
F[y,y']&=&\int \chi_\alpha(x_{0})\chi^*_\alpha(x'_{0}) dx_{0}dx'_0 dx_{1}\nn\\ 
&&\hspace{-1cm} \times \left\{ 1 + \frac{i\lambda\hbar}{2} \int_{0}^T dt \left( y_t^2\frac{\delta^2}{\delta f(t)^2}-y'^2_t\frac{\delta^2}{\delta f'(t)^2}\right)\right. \nn \\ 
&&\hspace{-1cm} \left.- \frac{\lambda^2\hbar^2}{4}  \int_{0}^T \int_{0}^t dt ds \left(y^2_t\frac{\delta^2}{\delta f(t)^2}-y'^2_t\frac{\delta^2}{\delta f'(t)^2}\right)\right.\nn\\
&&\hspace{+2cm}\left.\left(y^2_s\frac{\delta^2}{\delta f(s)^2}-y'^2_s\frac{\delta^2}{\delta f'(s)^2}\right) \right\} \nn \\ 
&&\hspace{-1cm} K_f(x_1,T;x_0,0)K_{f'}^*(x_1,T;x_0',0)|_{f=f'=0},
\label{eq:influencefuncbis}
\eea
where $f$, $f'$ are arbitrary functions and in particular $f'$ is distinct from the time derivative of $f$.

In order to compute this quantity, we notice that the three standard integrals over $x_0$, $x_0'$, and $x_1$ commute with the functional derivative operators, and therefore the above expression reduces to
\be
F[y,y']=\mathcal{D}[y,y']\mathcal{F}[f,f']|_{f=f'=0},
\ee
where the functional differential operator $\mathcal{D}[y,y']$ is given by
\bea
\mathcal{D}[y,y'] &=& 1 + \frac{i\lambda\hbar}{2} \int_{0}^T dt \left( y_t^2\frac{\delta^2}{\delta f(t)^2}-y'^2_t\frac{\delta^2}{\delta f'(t)^2}\right) \nn \\ 
&&\hspace{-1.5cm}- \frac{\lambda^2\hbar^2}{4}  \int_{0}^T \int_{0}^t dt ds \left(y^2_t\frac{\delta^2}{\delta f(t)^2}-y'^2_t\frac{\delta^2}{\delta f'(t)^2}\right)\nn\\
&&\hspace{+1.5cm}\left(y^2_s\frac{\delta^2}{\delta f(s)^2}-y'^2_s\frac{\delta^2}{\delta f'(s)^2}\right),
\eea
and
\bea
\mathcal{F}[f,f']&=&\int \chi_\alpha(x_{0})\chi^*_\alpha(x'_{0}) dx_{0}dx'_0 dx_{1}\nn\\
&&\hspace{0.3cm}\times K_f(x_1,T;x_0,0)K_{f'}^*(x_1,T;x_0',0).
\eea
Now $\mathcal{F}[f,f']$ can  be further simplified by realizing that
\be
K_f(x_1,T;x_0,0)=\bra{x_1}\hat{U}_f(T,0)\ket{x_0},
\ee
where $\hat{U}_f(T,0)\equiv\mathcal{T}e^{-\frac{i}{\hbar}\int_0^Tdt (\hat{H}_0(M,\Omega)-f(t)\hat{x})}$, $\hat{H}_0(M,\Omega)\equiv \frac{\hat{p}_x^2}{2M}+\frac{1}{2}M\Omega^2\hat{x}^2$ is the Hamiltonian operator derived from $L_0(M,\Omega;x]$, and $\mathcal{T}$ is the time ordering operator (with an analogous expression for $K_{f'}(x_1,T;x'_0,0)$). We thus obtain
\be
\mathcal{F}[f,f']=\bra{\alpha}\hat{U}_{f'}(T,0)^\dag\hat{U}_{f}(T,0)\ket{\alpha}.
\ee
Moreover, it can be shown that $\hat{U}_{f}(T,0)=e^{-\frac{i}{\hbar}\hat{H}_0(M,\Omega)T}\mathcal{T}e^{\frac{i}{\hbar}\int_0^Tdt f(t)\hat{x}_I(t)}$, where the interaction picture operator $\hat{x}_I(t)\equiv e^{+\frac{i}{\hbar}\hat{H}_0(M,\Omega)t}\hat{x}e^{-\frac{i}{\hbar}\hat{H}_0(M,\Omega)t}=\sqrt{\frac{\hbar}{2M\Omega}}(\hat{a}e^{-i\Omega t}+\hat{a}^\dag e^{+i\Omega t})$, and therefore the previous expression simply reduces to
\be
\mathcal{F}[f,f']=\bra{\alpha}\bar{\mathcal{T}}e^{-\frac{i}{\hbar}\int_0^Tdt f(t)\hat{x}_I(t)}\mathcal{T}e^{+\frac{i}{\hbar}\int_0^Tdt f(t)\hat{x}_I(t)}\ket{\alpha},
\ee
where $\bar{\cal T}$ is the anti-time-ordering operator.

In this form it is straightforward (albeit somewhat tedious) to compute $\mathcal{F}$ explicitly. First of all, since $[\hat{a},\hat{a}^\dag]=1$ is a c-number (and therefore commutes with every operator), the Magnus formula \cite{BLANES2009151} allows us the get rid of the time ordering operators at the expense of a multiplicative factor. Second, we can use the Baker-Campbell-Hausdorff formula \cite{Sakurai:1167961} to arrange the leftover ladder operators in normal order (up to another multiplicative factor). Lastly, we can use the fact that the coherent state $\ket{\alpha}$ is by definition an eigenstate of the annihilation operator and verifies $\hat{a}\ket{\alpha}=\alpha\ket{\alpha}=X_0\sqrt{M\Omega/2\hbar}\ket{\alpha}$. We thus obtain
\be
\mathcal{F}[f,f']=\exp i\Theta[f,f'],
\ee
where 
\begin{widetext}
\bea
i\Theta[f,f']=\frac{i}{\hbar}\int_0^T dt X_0\cos\Omega t\, (f(t)-f'(t))-\frac{1}{2M\hbar\Omega}\int_0^T\int_0^t dt ds\,(f(t)-f'(t))\left(f(s)e^{-i\Omega(t-s)}-f'(s)e^{i\Omega(t-s)}\right)
\label{eq:genfunc}
\eea
Notice that this expression coincides with the influence functional encoding the effect of a single harmonic oscillator at zero temperature as well as that of a classical time-dependent quadratic potential on the degree of freedom represented by $f(t)$ \cite{FEYNMAN1963118}.

With this equation at hand, applying the differential operator $\mathcal{D}[y,y']$ to $\mathcal{F}$ and setting $f=f'=0$ is once again just a matter of tedious bookkeeping and one finally obtains 
\bea
F[y,y']&=&1- \frac{i\lambda X_0^2}{2\hbar} \int_{0}^T dt \cos^2\Omega t\,(y^2_t-y'^2_t)  - \frac{i\lambda}{4M\Omega} \int_{0}^T dt(y^2_t-y'^2_t)\nn\\
&&\hspace{-2cm}-\frac{\lambda^2}{4}  \int_{0}^T \int_{0}^t dt ds\left[\frac{X_0^4}{\hbar^2} \cos^2{\Omega t}\cos^2{\Omega s}+\frac{X_0^2}{2M\hbar\Omega}(\cos^2{\Omega t}+\cos^2{\Omega s})+\frac{1}{4M^2\Omega^2}\right] (y^2_t-y'^2_t)(y^2_s-y'^2_s)\label{eq:IFfull} \\
&&\hspace{-2cm}-\frac{\lambda^2}{4}  \int_{0}^T \int_{0}^t dt ds\, (y^2_t-y'^2_t)\left[\frac{2X_0^2}{M\hbar\Omega}\cos{\Omega t}\cos{\Omega s}\left(y^2_s e^{-i\Omega(t-s)}-y'^2_s e^{i\Omega(t-s)}\right)+ \frac{1}{2M^2\Omega^2}\left(y^2_s e^{-2i\Omega(t-s)} - y'^2_s e^{2i\Omega(t-s)} \right)\right].\nn
\eea
\end{widetext}
Although this is the exact result, we can rewrite it in a more compact form by noticing that the expression takes the form $F[y,y'] = 1+\lambda a +\lambda^2(b+a^2/2)\approx e^{\lambda a + \lambda^2 b}$ at this order in perturbation theory, where $a$ and $b$ can be read off from the first and third line of the above equation respectively. We can then rewrite~\eqref{eq:IFfull} as
\be
F[y,y']=\exp i\Phi[y,y'],
\ee
where
\begin{widetext}
\bea
i\Phi[y,y']&=&-\frac{i\lambda X_0^2}{2\hbar} \int_{0}^T dt \cos^2\Omega t\,(y^2_t-y'^2_t)-\frac{i\lambda}{4M\Omega} \int_{0}^T dt\,(y^2_t-y'^2_t)\label{eq:InfluencePhase}\\
&&\hspace{-2cm}-\frac{\lambda^2}{4}  \int_{0}^T \int_{0}^t dt ds \, (y^2_t-y'^2_t)\left[\frac{2X_0^2}{M\hbar\Omega}\cos{\Omega t}\cos{\Omega s}\left(y^2_s e^{-i\Omega(t-s)}-y'^2_s e^{i\Omega(t-s)}\right)+ \frac{1}{2M^2\Omega^2}\left(y^2_s e^{-2i\Omega(t-s)} - y'^2_s e^{2i\Omega(t-s)} \right)\right].\nn
\eea
\end{widetext}
In the next section, we shall exploit this influence functional to better understand the quantum dynamics of the $y$ degree of freedom.

\section{Parametric resonance and squeezing}
\label{sec:eff}

Going back to Eq.~\eqref{eq:reduced}, we can use the expression for the influence phase~\eqref{eq:InfluencePhase} to rewrite the reduced density matrix element as
\bea
\bra{y_1}\hat{\rho}_R(T)\ket{y_2}&=&\int {\cal D}y{\cal D}y' e^{\frac{i}{\hbar}\left(\tilde{S}_0(m;\omega_{\rm eff},y]-\tilde{S}_0(m;\omega_{\rm eff},y']\right)}\nn\\
&&\hspace{-2.5cm}\times\left\{1-\frac{\lambda^2}{4}  \int_{0}^T \int_{0}^t dt ds \, (y^2_t-y'^2_t)\right.\nn\\
&&\hspace{-2cm}\times\left[\frac{2X_0^2}{M\hbar\Omega}\cos{\Omega t}\cos{\Omega s}\left(y^2_s e^{-i\Omega(t-s)}-y'^2_s e^{i\Omega(t-s)}\right)\right.\nn\\&&\hspace{-0.8cm}\left.\left.+ \frac{1}{2M^2\Omega^2}\left(y^2_s e^{-2i\Omega(t-s)} - y'^2_s e^{2i\Omega(t-s)} \right)\right]\right\}\nn\\
&&\hspace{-2.5cm}\times \psi_0(y_0) \psi_0^*(y_0') dy_{0} dy'_{0},
\label{eq:densitymatcomp}
\eea
where we remind the reader that the boundary conditions on the path integral are $y(0)=y_0$, $y'(0)=y'_0$, $y(T)=y_1$, and $y'(T)=y_2$.
In the first line, we absorbed the terms linear in $\lambda$ in the influence phase~\eqref{eq:InfluencePhase} into a free action on an explicitly time-dependent background that reads
\be
\tilde{S}_0(m;\omega_{\rm eff},y]=\int_0^T dt\left(\frac{1}{2}m\dot{y}^2 -\frac{1}{2}m\omega^2_{\mathrm{eff}}(t)y^2\right),
\ee
where 
\bea
\omega_{\rm eff}^2(t)&=&\omega^2+\frac{\lambda \hbar}{2mM\Omega}+\frac{\lambda X_0^2}{m}\cos^2{\Omega t}\nn\\&=& \omega_r^2-\frac{\lambda X_0^2}{m}\sin^2{\Omega t},
\label{eq:paramosc}
\eea
and 
\be
\omega_{r}^2\equiv \omega_{\mathrm{eff}}^2(0)=\omega^2+\frac{\lambda \hbar}{2mM\Omega}+\frac{\lambda X_0^2}{m}
\label{eq:renfreq}
\ee
is a {\it renormalized angular frequency} for the $y$ oscillator. As expected, given the choice of quantum state for the $x$ oscillator, this is the action for a parametric oscillator whose angular frequency is time dependent but periodic with period $\pi/\Omega$. Notice that~\eqref{eq:paramosc} is not the standard way of parametrizing such systems, and in fact we will find it useful to also introduce the more standard parametrization
\be
\omega_{\rm eff}^2(t)=\omega'^2\left(1+2\delta\cos 2\Omega t\right),
\ee
where
\be
\omega'^2\equiv \omega^2+\frac{\lambda \hbar}{2mM\Omega}+\frac{\lambda X_0^2}{2m}=\omega_r^2-\frac{\lambda X_0^2}{2m},
\label{eq:omegaprime}
\ee
and
\be
\delta \equiv \frac{\lambda X_0^2}{4m\omega'^2}.
\ee
The remaining part of the IF, corresponding to the quadratic terms in $\lambda$ in the influence phase~\eqref{eq:InfluencePhase}, was Taylor expanded to lowest non-trivial order yielding the terms in curly brackets. 

Before going any further it is necessary to specify the  state $\ket{0}$ or in other words the wavefunction $\psi_0$. Since the degree of freedom $y$ evolves in a time dependent background, there exists no ground state in the usual sense of a stationary state with minimal energy. However, one can define a non-unique notion of {\it adiabatic ground state} \cite{Birrell_Davies_1982} and in particular a {\it zeroth-order ground state} as the instantaneous ground state of the frozen system at $t=0$. This corresponds to the ground state of a simple harmonic oscillator described by the Hamiltonian $\hat{H}_0(m,\omega_r)$. 

It is easy to see that the $\mathcal{O}(\lambda^2)$ terms in Eq.~\eqref{eq:densitymatcomp} are responsible for entanglement between the $x$ and $y$ systems and we shall explore this more quantitatively in the next section. At this level though, it is sufficient to observe that neglecting these terms leads to $\hat{\rho}_R(T)=\ket{\psi(T)}\bra{\psi(T)}$ where $\bra{\psi(T)}$ is obtained from $\bra{0}$ by the action of the unitary evolution operator associated to the explicitly time-dependent Hamiltonian operator $\hat{H}_0(m,\omega_{\rm eff}(t))$.

To proceed further in the evaluation of the reduced density matrix elements, we use the functional methods presented
in the previous section and introduce the propagation kernel
\be
\mathcal{K}_f(y_b,t_b;y_a,t_a)\equiv
\int {\cal D}y\, e^{\frac{i}{\hbar}\int_{t_a}^{t_b}dt\left(\frac{1}{2}m\dot{y}^2-\frac{1}{2}m\omega_{\rm eff}^2(t) y^2+fy\right)},
\ee
where the boundary conditions are $y(t_a)=y_a$ and $y(t_b)=y_b$. Analogously to~\eqref{eq:influencefuncbis}, we can then rewrite the reduced density matrix element as
\bea
\bra{y_1}\hat{\rho}_R(T)\ket{y_2}&=&\int \psi_0(y_0) \psi_0^*(y_0') dy_{0} dy'_{0}\nn\\
&&\hspace{-3cm}\times\left\{1-\frac{\lambda^2\hbar^4}{4}  \int_{0}^T \int_{0}^t dt ds \, 
\left(\frac{\delta^2}{\delta f(t)^2}-\frac{\delta^2}{\delta f'(t)^2}\right)\right.\nn\\
&&\hspace{-2.5cm}\times\left[\frac{2X_0^2}{M\hbar\Omega}\cos{\Omega t}\cos{\Omega s}\right.\nn\\&&\hspace{-1cm}\times\left(\frac{\delta^2}{\delta f(s)^2} e^{-i\Omega(t-s)}-\frac{\delta^2}{\delta f'(s)^2} e^{i\Omega(t-s)}\right)\nn\\&&\hspace{-2.5cm}\left.\left.+ \frac{1}{2M^2\Omega^2}\left(\frac{\delta^2}{\delta f(s)^2} e^{-2i\Omega(t-s)} - \frac{\delta^2}{\delta f'(s)^2} e^{2i\Omega(t-s)} \right)\right]\right\}\nn\\
&&\hspace{-2.5cm}\times\ \mathcal{K}_f(y_1,T;y_0,0)\mathcal{K}_{f'}^*(y_2,T;y_0',0)|_{f=f'=0}.
\label{eq:densitymatcompbis}
\eea

As before, we can express the propagation kernels in canonical formalism as 
\be
\mathcal{K}_f(y_1,T;y_0,0)=\bra{y_1}\hat{\cal U}_f(T,0)\ket{y_0},
\ee
where $\hat{\cal U}_f(T,0)\equiv\mathcal{T}e^{-\frac{i}{\hbar}\int_0^Tdt (\hat{H}_0(m,\omega_{\rm eff}(t))-f(t)\hat{y})}$. Contrary to~\eqref{eq:influencefuncbis}, there is no integration over the variables $y_1$ and $y_2$, so Eq.~\eqref{eq:densitymatcompbis} cannot be further simplified at this point. 

However, we can use it to compute all the moments of the marginal probability density for $y$ at time $T$---in other words, all expectation values of the form ${\rm Tr}[\hat{\rho}_R(T)\hat{y}^n]$. This is achieved by setting $y_1=y_2$ in~\eqref{eq:densitymatcompbis}, integrating over it, and taking an appropriate number of functional derivatives as in\footnote{In fact this trick can be adapted to compute any number of Heisenberg picture correlation functions of the type $\langle\hat{y}(t'_1)\dots\hat{y}(t'_n)\hat{y}(t_{m})\dots\hat{y}(t_{1})\rangle \equiv {\rm Tr}\left[\hat{\rho}_R(0)\hat{y}(t'_1)\dots\hat{y}(t'_n)\hat{y}(t_{m})\dots\hat{y}(t_{1})\right]$, where $0\leq t_1\leq t_2\leq \dots \leq t_n\leq T$ and $0\leq t'_1\leq t'_2\leq \dots \leq t'_m\leq T$ so that the $n$ leftmost position operators are anti-time-ordered and the $m$ rightmost ones are time-ordered.}
\be
{\rm Tr}[\hat{\rho}_R(T)\hat{y}^n] = \left(\frac{\hbar}{i}\frac{\delta}{\delta f(T)}\right)^n \mathcal{D}'\, \mathcal{G}[f,f']|_{f=f'=0},
\label{eq:moments}
\ee
where the differential operator $\mathcal{D}'$ is defined by
\bea
\mathcal{D}'&=&1-\frac{\lambda^2\hbar^4}{4}  \int_{0}^T \int_{0}^t dt ds \, 
\left(\frac{\delta^2}{\delta f(t)^2}-\frac{\delta^2}{\delta f'(t)^2}\right)\nn\\
&&\hspace{-1cm}\times\left[\frac{2X_0^2}{M\hbar\Omega}\cos{\Omega t}\cos{\Omega s}\right.\label{eq:diffoperator}\\&&\ \ \quad\times\left(\frac{\delta^2}{\delta f(s)^2} e^{-i\Omega(t-s)}-\frac{\delta^2}{\delta f'(s)^2} e^{i\Omega(t-s)}\right)\nn\\&&\hspace{-1cm}\quad\left.+ \frac{1}{2M^2\Omega^2}\left(\frac{\delta^2}{\delta f(s)^2} e^{-2i\Omega(t-s)} - \frac{\delta^2}{\delta f'(s)^2} e^{2i\Omega(t-s)} \right)\right],\nn
\eea
and 
\be
\mathcal{G}[f,f']=\bra{0}\hat{\cal U}_{f'}(T,0)^\dag\hat{\cal U}_{f}(T,0)\ket{0}.
\label{eq:genfunc}
\ee

Analogously to the previous section, we can compute $\mathcal{G}[f,f']$ explicitly by noticing that $\hat{\cal U}_{f}(T,0)=\mathcal{T}e^{-\frac{i}{\hbar}\int_0^T dt \hat{H}_0(m,\omega_{\rm eff}(t))}\mathcal{T}e^{\frac{i}{\hbar}\int_0^T dt f(t)\hat{y}_I(t)}$, where this time the interaction picture operator is defined with respect to the time-dependent Hamiltonian $\hat{H}_0(m,\omega_{\rm eff}(t))$ by 
\bea
\hat{y}_I(t)&\equiv&\bar{\mathcal{T}}e^{+\frac{i}{\hbar}\int_0^T dt \hat{H}_0(m,\omega_{\rm eff}(t))}\hat{y}\mathcal{T}e^{-\frac{i}{\hbar}\int_0^T dt \hat{H}_0(m,\omega_{\rm eff}(t))}\nn\\
&=&\sqrt{\frac{\hbar}{2m\omega_r}}\left(z(t)\hat{a}+z^*(t)\hat{a}^\dag \right).
\eea
Here the complex mode function $z(t)$ verifies
\be
\ddot{z}+\omega_{\rm eff}^2(t)z=0,
\label{eq:Mathieu}
\ee
with initial conditions $z(0)=1$ and $\dot{z}(0)=-i\omega_r$. When the coupling $\lambda$ vanishes, it reduces to the familiar $e^{-i\omega t}$.

With these notations at hand the functional $\mathcal{G}[f,f']$ reads
\be
\mathcal{G}[f,f']=\bra{0}\bar{\mathcal{T}}e^{-\frac{i}{\hbar}\int_0^Tdt f(t)\hat{y}_I(t)}\mathcal{T}e^{+\frac{i}{\hbar}\int_0^Tdt f(t)\hat{y}_I(t)}\ket{0},
\ee
and we can immediately infer the final result from Eq.~\eqref{eq:genfunc} by setting $\alpha$ to zero and doing the replacements $M\to m$, $\Omega \to \omega_r$, $e^{-i\Omega t}\to z(t)$. We obtain
\be
\mathcal{G}[f,f']\equiv \exp i\vartheta[f,f'],
\label{eq:genfuncG}
\ee
where
\bea
i\vartheta[f,f']&=&-\frac{1}{2m\hbar\omega_r}\int_0^T\int_0^t dt ds\,(f(t)-f'(t))\nn\\
&&\ \times\left(f(s)z(t)z^*(s)-f'(s)z^*(t)z(s)\right).
\label{eq:genfunctheta}
\eea

We can now use this expression along with Eq.~\eqref{eq:moments} to compute the statistical moments of $\hat{y}$. The $n=1$ case yields the expectation value of $\hat{y}$, and it is clear that this will vanish for all times given that $\vartheta[f,f']$ is quadratic in $f$ and $f'$. The $n=2$ case yields the variance of $\hat{y}$ and this can also be easily computed. In particular, when we neglect terms of order $\mathcal{O}(\lambda^2)$ in $\mathcal{D}'$, {\it i.e.} when we neglect quantum entanglement between $\hat{x}$ and $\hat{y}$, the result is
\be
{\rm Tr}[\hat{\rho}_R(T)\hat{y}^2] = -\hbar^2\frac{\delta^2}{\delta f(T)^2} \mathcal{G}[f,0]|_{f=0}
=\frac{\hbar}{2m\omega_r}|z(T)|^2.
\ee
The state therefore appears to be {\it squeezed} by a factor of $|z(T)|^2$, that is, its fluctuations are reduced/enhanced with respect to the chosen vacuum state.

We have already seen that $z(t)$ obeys Eq.~\eqref{eq:Mathieu}, which is the complexified version of the classical Eq.~\eqref{eq:classicalMathieu} (up to $\hbar$-dependent corrections). This differential equation is of the Mathieu type and, as mentioned in  the introduction, it is the paradigmatic example of a dynamical system that may, depending on the parameters, exhibit {\it parametric resonance}. This is a phenomenon whereby an oscillator with periodically-varying parameters will undergo an exponential enhancement of the amplitude of oscillation at certain resonant frequencies (see e.g. \cite{landau1976mechanics, McLachlan_1951}). In the case of the Mathieu equation~\eqref{eq:Mathieu}, parametric resonance is known to occur for values of the parameters in the so-called {\it instability bands} of the full parameter space. In these bands, the amplitude of oscillation grows exponentially at a rate set by the Floquet exponents of the system, and more precisely by the Floquet exponent with the largest positive real part (i.e. Lyapunov exponent); outside the resonance bands, all Floquet exponents are purely imaginary and the system is stable. Using the machinery of Floquet theory \cite{landau1976mechanics}, the instability bands can be shown to be approximately centered on the surfaces of equation 
\begin{equation}
    \omega'=n\Omega,
    \label{eq:resbandrelation}
\end{equation}
where $n$ is a positive integer. Therefore, when our parameters $\omega,\,\lambda,\,X_0,\,\dots$ are close to this surface, the system is expected to be parametrically unstable so that $|z(t)|$ grows without bounds. We thus conclude that parametric resonance leads to runaway squeezing of the quantum state of $\hat{y}$. Of course, this is only strictly valid while backreaction is negligible and eventually the growth of $|z(t)|$ will saturate, but this requires considering higher order in $\lambda$ corrections such as the ones we neglected when approximating $\mathcal{D}'$ by 1.

As we briefly alluded to above, quantum entanglement  only manifests when including order $\mathcal{O}(\lambda^2)$ terms in $\mathcal{D}'$. In the next section, we shall see that in this case as well parametric resonance plays an important role.

\section{Parametric resonance and entanglement}
\label{sec:ent}

We have at our disposal an expression for the reduced density matrix~\eqref{eq:densitymatcompbis} at time $T$, which should allow us to determine to what extent the quantum state of the system $y$ is {\it mixed}. This statistical mixing is due to the quantum entanglement between systems $x$ and $y$ generated during time evolution by the interaction term $\lambda x^2 y^2$. There exist many measures of entanglement/mixing but probably the most popular of all is the {\it entanglement entropy} defined as the von Neumann entropy of the reduced density matrix,
\be
S_E\equiv -{\rm Tr}\left[\hat{\rho}_R\ln{\hat{\rho}_R}\right],
\ee
which has a nice information theoretic interpretation \cite{Horodecki_2009}. Here and henceforth we have dropped the explicit time dependence of the density matrix $\hat{\rho}_R(T)$, implicitly assuming that all occurrences of quantities whose time argument is missing are taken at time $T$.  Another measure is the {\it purity}, $\gamma$, defined as the trace of the square of the reduced density matrix,
\be
\gamma \equiv {\Tr}\left[\hat{\rho}_R^2\right].
\ee
For an infinite-dimensional Hilbert space, the purity takes values between 0 and 1, and is particularly well-adapted to determining whether a state is pure or mixed: $\gamma=1$ if and only if the system is in a pure state, while $\gamma=0$ represents a maximally mixed state.

Conveniently, both the purity and entanglement entropy can be calculated from the {\it R\'enyi entropy}, defined as
\begin{equation}
    S_n=\frac{1}{1-n}\ln\mathrm{Tr}\left[\hat{\rho}_R^n\right],
    \label{eq:renyi},
\end{equation}
by noticing that $S_2=-\ln{\gamma}$ and $S_1\equiv\lim_{n\to 1} S_n=S_E$. Using Eqs.~\eqref{eq:densitymatcompbis},~\eqref{eq:diffoperator} and~\eqref{eq:genfunc}, while also adding a lower index to the operator $\mathcal{D}'$ to keep track of the functional variables it acts upon ({\it e.g.} $\mathcal{D}'_i$ will only contain functional derivatives with respect to $f_i$ and $f_i'$), we find 
\bea
    \mathrm{Tr}\left[\hat{\rho}_R^n\right]&=&\int dy_1\dots dy_n\bra{y_1}\hat{\rho}_R\ket{y_2}\bra{y_2}\hat{\rho}_R\ket{y_3}\dots\nn\\
&&\hspace{+2cm}\dots\bra{y_{n-1}}\hat{\rho}_R\ket{y_n}\bra{y_n}\hat{\rho}_R\ket{y_1}\nn\\
    && \hspace{-1.5cm}=\mathcal{D}'_{1}\mathcal{D}'_{2}\dots\mathcal{D}_n'\mathcal{G}[f_2,f_1']\mathcal{G}[f_3,f_2']\dots\mathcal{G}[f_1,f_n']|_{f_i=f_i'=0}.\nn
\eea
Noticing that $\mathcal{D}'_i=1+\lambda^2 \bar{\mathcal{D}}_i'$ (where $\bar{\mathcal{D}}_i'$ can be read off from Eq.~\eqref{eq:diffoperator}), the above expression reduces to
\bea
\mathrm{Tr}\left[\hat{\rho}_R^n\right]&=&1+n\lambda^2(\bar{\mathcal{D}}'_{1}+\bar{\mathcal{D}}'_{2})\mathcal{G}[f_2,f_1']\mathcal{G}[f_1,f_2']|_{f_i=f_i'=0}\nn\\
&=&1-\frac{n}{2}(1-\gamma).
\eea
up to terms of higher than second  order in $\lambda$. Here, we expressed the result in terms of  the $\mathcal{O}(\lambda^2)$ quantity $1-\gamma$ obtained from setting $n=2$ in the formula.

Thus, at this order in perturbation theory it is sufficient to compute the purity and all the other entropies can generically be obtained from it. The only caveat to this statement is the $n=1$ case corresponding to the entanglement entropy. Even though the above computation is only strictly valid for $n\geq 2$, we could in principle analytically continue the result and take the limit $n\to 1$ in~\eqref{eq:renyi}. We would then obtain
\be
S_n\sim -\frac{n(1-\gamma)}{2(1-n)},
\ee
which is singular as $n\to 1$. This may seem puzzling at first glance, but this is simply a sign of the fact that the von Neumann entropy $S_E$ of the subsystem $y$ is non-analytic at $\lambda=0$. This is to be expected since $\hat{\rho}_R$ would be a projector matrix for all times if the coupling were turned off, with one unit eigenvalue and the rest being zeros. This implies that the matrix $\hat{\rho}_R\ln\hat{\rho}_R$ is zero in this limit, but as $\lambda$ increases its eigenvalues become non-zero, making it scale as $\sim\lambda^2\ln{\lambda^2}$~\cite{Dadras:2020xfl}. A similar non-analytic behavior in the $T$ variable is also expected to occur around the initial time (for finite $\lambda$).

We will therefore concentrate our efforts on computing the purity, which should be a perfectly good measure of the amount of entanglement between the two quantum oscillators $x$ and $y$~\cite{Schneeloch:2020zkv}. Using the explicit formula for $\mathcal{G}[f,f']$ (see Eqs.~\eqref{eq:genfuncG} and~\eqref{eq:genfunctheta}), we obtain eventually\footnote{It will be convenient for us to present our quantitative results in the form $1-\gamma=\cdots$ since the zeroth order term in the purity $\gamma$ is always 1 and the next order correction is always negative.}
\be
1-\gamma = \frac{\hbar\lambda^2X_0^2}{2 m^2\omega_r^2 M\Omega}|I_1|^2+\frac{\hbar^2\lambda^2}{8m^2\omega_r^2M^2\Omega^2}|I_2|^2,
\label{eq:puritysimpl}
\ee
where
\bea
 I_1&=&\int_0^T dt \cos{\Omega t}\ z(t)^2  e^{-i\Omega t}\label{eq:I1},\\ I_2&=&\int_0^T dt\, z(t)^2  e^{-2i\Omega t}.
\eea
Recall that the mode function $z(t)$ is obtained by solving~\eqref{eq:Mathieu} with initial conditions $z(0)=1$ and $\dot{z}(0)=-i\omega_r$. Explicitly it reads
\be
    z(t)=\frac{\mathcal{C}(a, q, \Omega t)}{\mathcal{C}(a, q, 0)}-\frac{i\omega_r}{\Omega} \frac{\mathcal{S}(a,q,\Omega t)}{\mathcal{S'}(a,q,0)},
\ee
where $a=(\omega_r^2-\lambda X_0^2/2m)/\Omega^2=\omega'^2/\Omega^2$, $q=-\lambda X_0^2/4m\Omega^2=-\delta(\omega'/\Omega)^2$, and $\mathcal{C}$, $\mathcal{S}$ are the Mathieu C and S functions \cite{Weisstein}. The primed version of these functions denotes their derivative with respect to the third argument. Notice that both $I_1$ and $I_2$ are expressed in terms of windowed Fourier transforms of $z(t)^2$.

At this point it may be edifying to recast the coefficients of Eq.~\eqref{eq:puritysimpl} in terms of the dimensionless coupling constant $\tilde{\lambda}\equiv\hbar\lambda/mM\Omega\omega_r^2$ and the average occupation number $N\equiv M\Omega X_0^2/2\hbar$ of the coherent state $\ket{\alpha}$ to obtain 
\be
1-\gamma=N\tilde{\lambda}^2\omega_r^2 |I_1|^2+\frac{1}{8}\tilde{\lambda}^2\omega_r^2 |I_2|^2.
\ee
It is clear from this expression that since the purity is always between 0 and 1, the perturbative expansion will only be valid in the regime where both $\tilde{\lambda}\ll 1$ and the so-called {\it collective coupling} $N\tilde{\lambda}\ll 1$. This is in fact the way to give meaning to the small $\lambda$ expansion of the IF in Sec.~\ref{sec:IF}.

This expression, despite being exact, is not very helpful since it involves special functions. However, when the parameters are such that the system undergoes parametric resonance, for instance when $\omega'\approx \Omega$ (first instability band), we can express the mode function in an approximate analytic form in the limit where\footnote{Recall that $\omega'^2=\omega_r^2(1-\frac{\lambda X_0^2}{2m\omega_r^2})=\omega_r^2\left(1-N\tilde{\lambda}\right)$ (see Eq.~\eqref{eq:omegaprime}).} $\delta= \lambda X_0^2/4m\omega'^2\approx N\tilde{\lambda}/2 \ll 1$ as
\be
    z(t)=\cosh{\mu t}\, \left[e^{-i\omega' t}+\mathcal{O}(\delta)\right]+i\sinh{\mu t}\,\left[e^{i\omega' t}+\mathcal{O}(\delta)\right]
    \label{eq:floquetapprox}
\ee
where $\mu \approx \delta \omega'/2 \approx \lambda X_0^2/8m\Omega$ is the Lyapunov exponent of the system and the $\mathcal{O}(\delta)$ terms include higher harmonics such as $e^{\pm 3i\omega't}$ and so on~\cite{landau1976mechanics}. The full expansion can be computed via multiscale analysis and is valid until times $t\gtrsim 1/\mu$ (see e.g.~\cite{strogatz2015nonlinear}).

Comparing this expression to the more standard expansion in terms of the $\lambda=0$ mode functions,
\be
z(t)= \alpha(t) e^{-i\omega_r t} + \beta(t) e^{+i\omega_r t},
\ee
we can immediately read off the value of the Bogoliubov coefficients,
\be
\alpha(t) \approx \cosh{\mu t}\,e^{+i\delta \Omega t}, \quad \beta(t) \approx i\sinh{\mu t}\,e^{-i\delta \Omega t}.
\ee 
It is then easy to see that the average occupation number of the $y$ oscillator is $n_y(t)\equiv |\beta(t)|^2=\sinh^2{\mu t}$. As expected, the Lyapunov exponent $\mu$ also sets the squeezing amplitude $|z(t)|^2=\cosh{2\mu t}-\sinh{2\mu t}\sin{2\omega't}$.

We can in fact see how the purity evolves as a function of squeezing by using (\ref{eq:floquetapprox}) in (\ref{eq:puritysimpl}). In addition to the assumption $\delta\sim N\tilde{\lambda}\ll 1$, we will take $N\gg 1$, corresponding to a large occupation number in $\ket{\alpha}$, or in other words, an approximately classical evolution of $x$. Since in this regime $N\tilde{\lambda}\gg \tilde{\lambda}$, the first integral in (\ref{eq:puritysimpl}) dominates over the second one and  we will keep only this term in our approximation. When we insert the expression for the mode function into (\ref{eq:I1}) and carry out the integration, we obtain three types of oscillating terms: terms whose amplitude grows at most linearly in the time $T$, as well as terms whose amplitude grows like $\cosh(2\mu T)$ and $\sinh(2\mu T)$. For times $T\gtrsim 1/\mu$ the latter terms are both of order $\sim2\sinh^2 \mu T$, and they dominate the expression for $I_1$. We thus obtain
\be
I_1\approx\frac{\sinh^2\mu T}{4\mu}\left[-1+2i+\mathcal{O}(\delta)\right], 
\ee
and consequently
\be
    1-\gamma\approx\frac{5\hbar\lambda^2 X_0^2}{32 m^2\omega_r^2 M \Omega}\frac{\sinh^4(\mu T)}{\mu^2}.
    \label{eq:puritylatetime}
\ee
Thus, in terms of the occupation number, the purity scales as $1-\gamma\sim n_y^2$. These results point to a deep connection between particle production, squeezing, and entanglement in situations where parametric resonance is operative. In this case, the purity decreases significantly (at times to the point where it becomes negative, thus signaling a breakdown of perturbation theory) as the two quantum systems become more entangled and the quantum state for $y$ becomes more mixed; at the same time the fluctuations of $y$ are enhanced as a consequence of runaway squeezing. In the next section, we shall discuss more precise numerical results.

\section{Results and Discussion}
\label{sec:disc}

\begin{figure*}[ht]
    \centering
    \subfigure[\label{fig:purity_omega0.5}]{\includegraphics[width=0.45\linewidth]{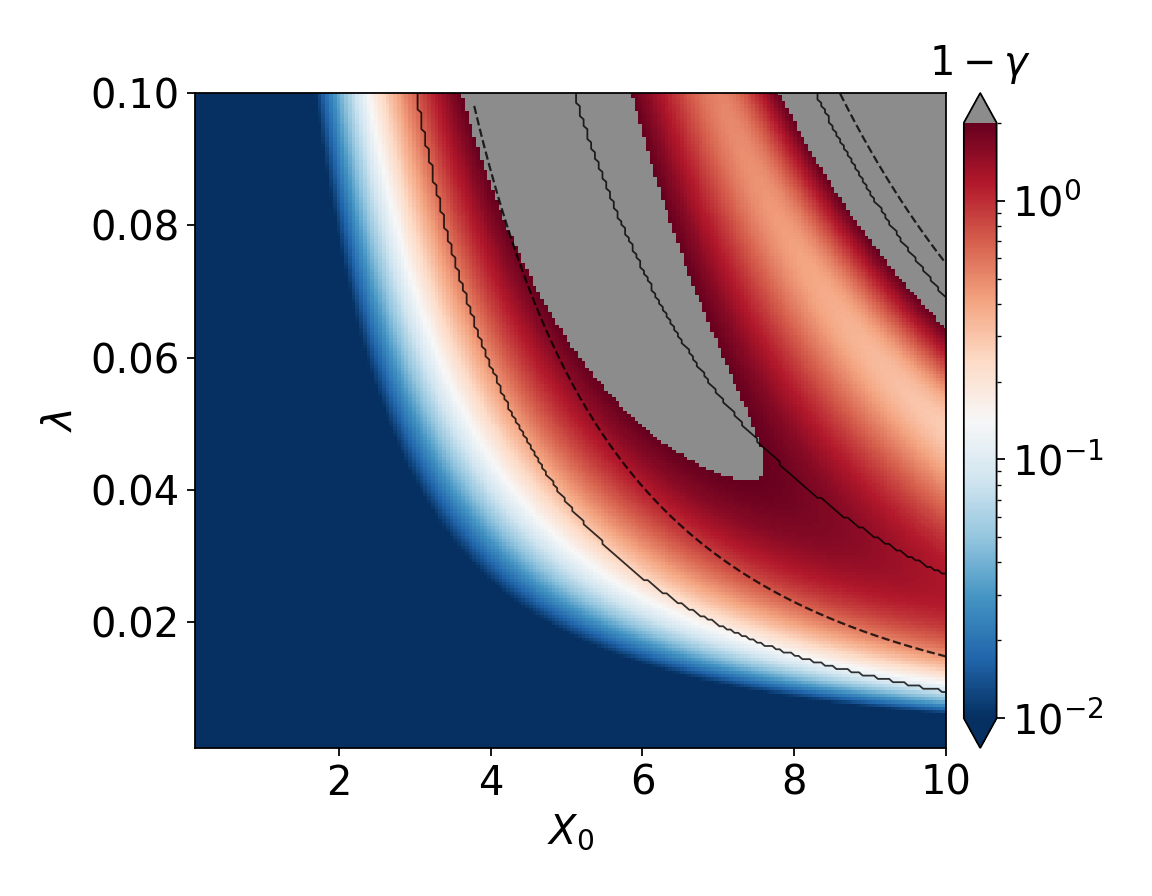}}\qquad
    \subfigure[\label{fig:purity_omega0.9}]{\includegraphics[width=0.45\linewidth]{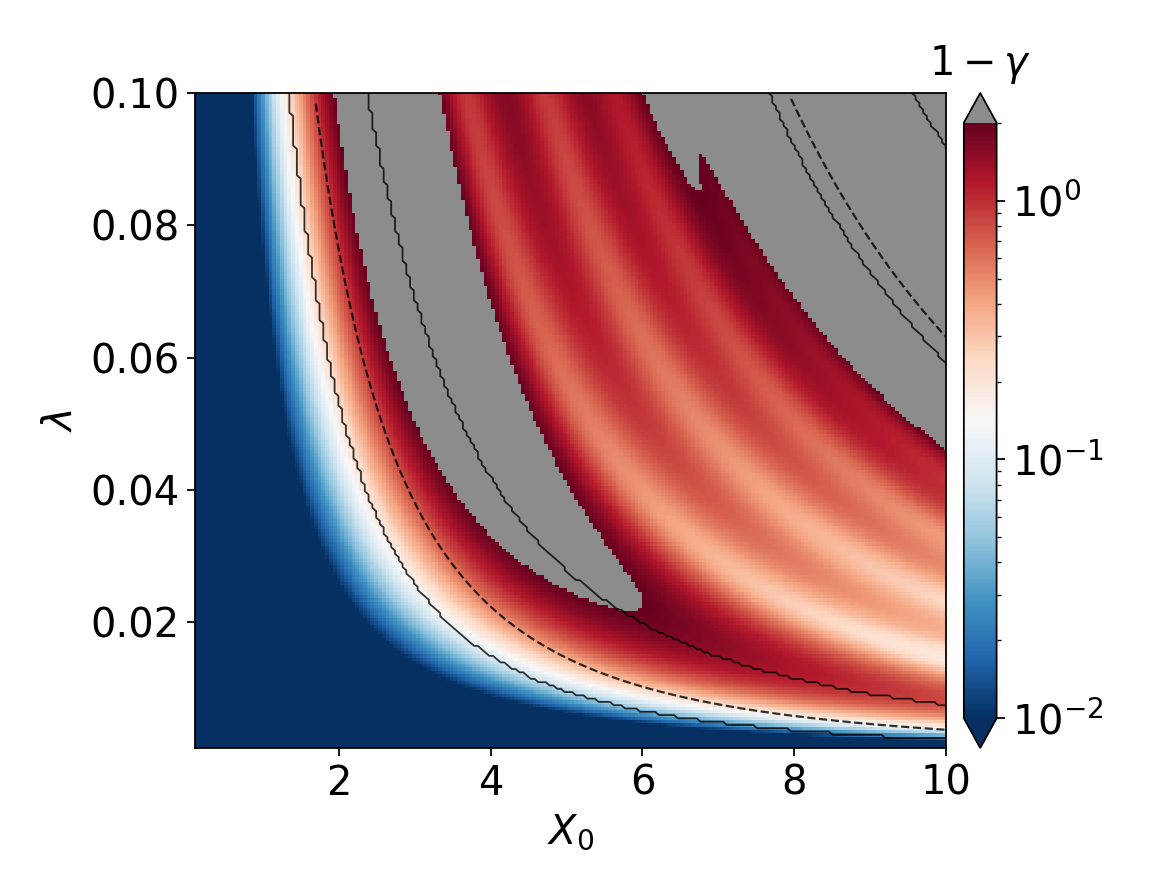}}
    \caption{Density plots  of $1-\gamma$ as a function of $X_0$ and $\lambda$ for (a) $\omega=0.5$ and $T=6$ and (b) $\omega=0.9$ and $T=20$. The center of the resonance bands is shown with a dotted line and the edges with a solid line. Gray regions correspond to values of $1-\gamma\gtrsim 1$.}
    \label{fig:purity_omega0.5_0.9}
\end{figure*}

\begin{figure*}[ht]
    \centering
    \subfigure[]{\includegraphics[width=0.45\linewidth]{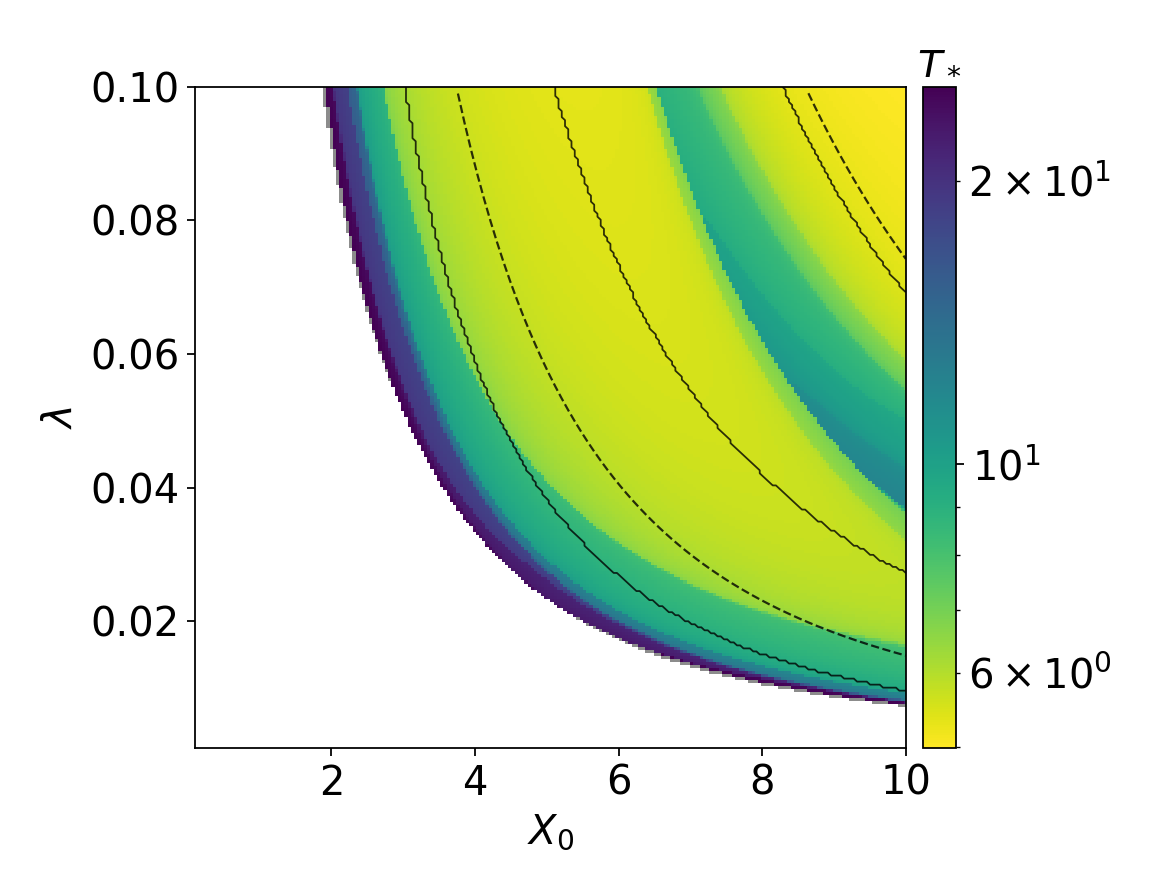}}\qquad
    \subfigure[]{\includegraphics[width=0.45\linewidth]{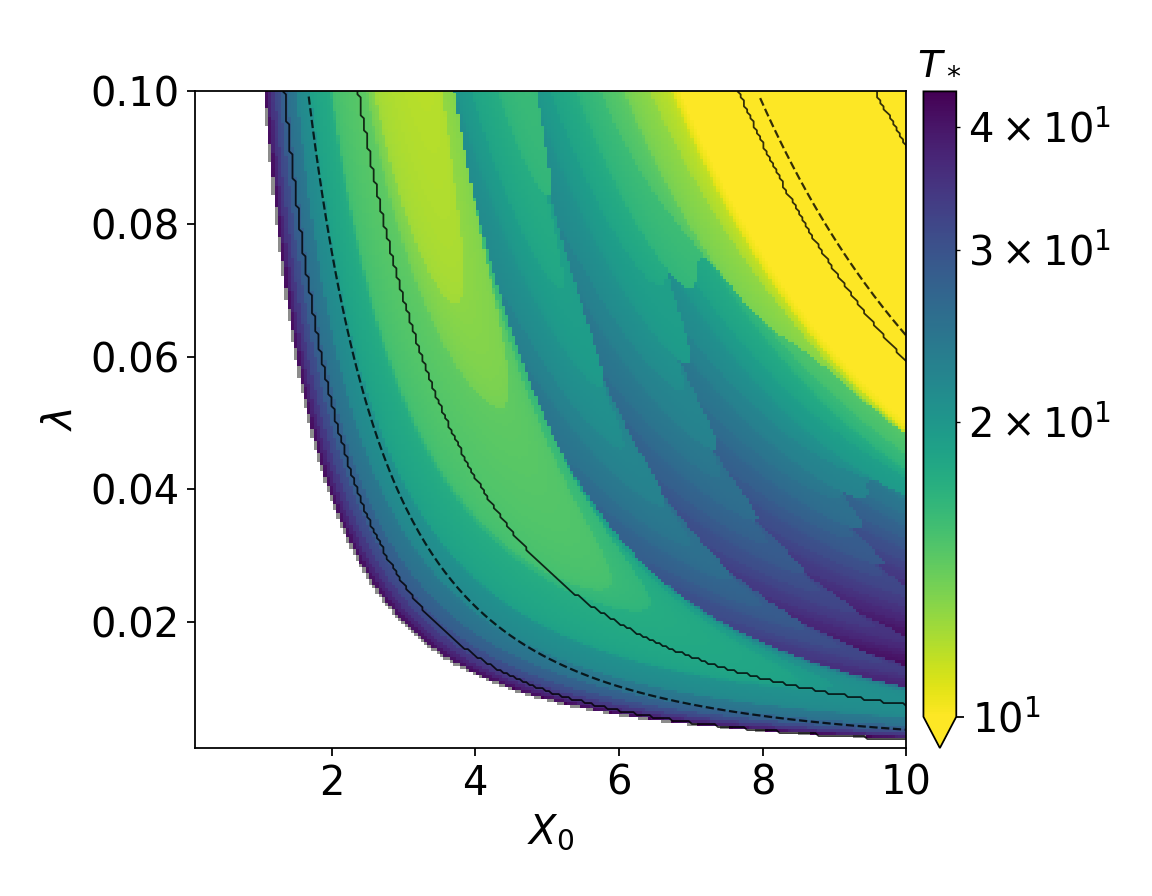}}
    \caption{Density plots showing the time $T_*$ at which $1-\gamma$ reaches 1 for (a) $\omega=0.5$ and (b) $\omega=0.9$, which is the time beyond which our perturbative expansion breaks down. For the non-colored part of the plot this time is not reached within (a) $T=25$ and (b) $T=45$. }
    \label{fig:cutoff_omega0.5_0.9}
\end{figure*}

For definitiveness we work in units where $\Omega=M=\hbar=1$. As mentioned before, a necessary condition for our perturbative expansion to remain valid is for the dimensionless quantity $\lambda/m\omega_r^2\sim\tilde{\lambda}$ to be much smaller than 1. $\lambda X_0^2/m\omega_r^2 \sim N\tilde{\lambda}$ is not assumed to be small although it will turn out to be the case for most regions of parameter space. 

For various values of $\omega$ we plot $1-\gamma$ (at some fixed time $T$) as a function of $\lambda$ and $X_0$. Recall that the instability bands are given by~\eqref{eq:resbandrelation}, which in our choice of units reduces to
\be
n^2=\omega^2 +\frac{\lambda}{2m}(X_0^2+1),
\ee
where $n=1,2,3\dots$. For the first resonance band ($n=1$) to be visible for positive values of $\lambda$, we need $\omega$ to be less than 1. In the following, we will pick $m=1$ and use the ranges $X_0\in[0.5,10]$, $\lambda\in[0.01,0.1]$.

Figure~\ref{fig:purity_omega0.5_0.9} shows the behavior of $1-\gamma$ for this range of $X_0$ and $\lambda$, and for two different values of $\omega$. We have chosen the time $T$ such that our perturbative approximation~\eqref{eq:puritysimpl} is mostly valid. Values of $1-\gamma$ which do exceed the threshold of $\approx 1$, i.e. for which the perturbative expansion breaks down, are denoted in gray. Clearly, the purity decreases more quickly in the resonance band, as indicated by the dark red/gray regions aligning closely with the bands computed directly from the Mathieu equation~(\ref{eq:Mathieu}). For values of $\omega$ farther from $1$, represented by $\omega=0.5$ in Fig.~\ref{fig:purity_omega0.5}, this trough in the purity settles more quickly on the band than for values closer to $1$, shown by $\omega=0.9$ in Fig.~\ref{fig:purity_omega0.9}, as can be seen from the times chosen for the plots (this is further examined in Fig.~\ref{fig:cutoff_omega0.5_0.9}). We consider our results to be trustworthy only for the first resonance band, even though the second band, appearing in the upper right corner of the plot, also shows a strong decrease in purity. This is due to this region being in the ``strong-collective-coupling'' regime, where $N\tilde{\lambda}\ll 1$ does not hold, so our perturbative result is unreliable.

Figure \ref{fig:cutoff_omega0.5_0.9} presents the range of validity for our perturbative expansion by showing the time $T_{*}$ at which $1-\gamma$ reaches the value 1. As expected, this threshold is reached much faster in the resonance band. Indeed, $T_*\approx 6$ for $\omega=0.5$ and $T_*\approx 10$ for $\omega=0.9$, coinciding with the trough in the purity in the previous figure. This quantity was computed only up to $T=25$ in the first case and $T=45$ in the second, as going beyond that did not provide more relevant information: indeed there is a region before the first resonance band (in the lower left corner) in which the purity stays positive on the considered time-scales and hence $T_*$ is effectively infinite.

\begin{figure}[h]
    \centering
    \includegraphics[width=0.9\linewidth]{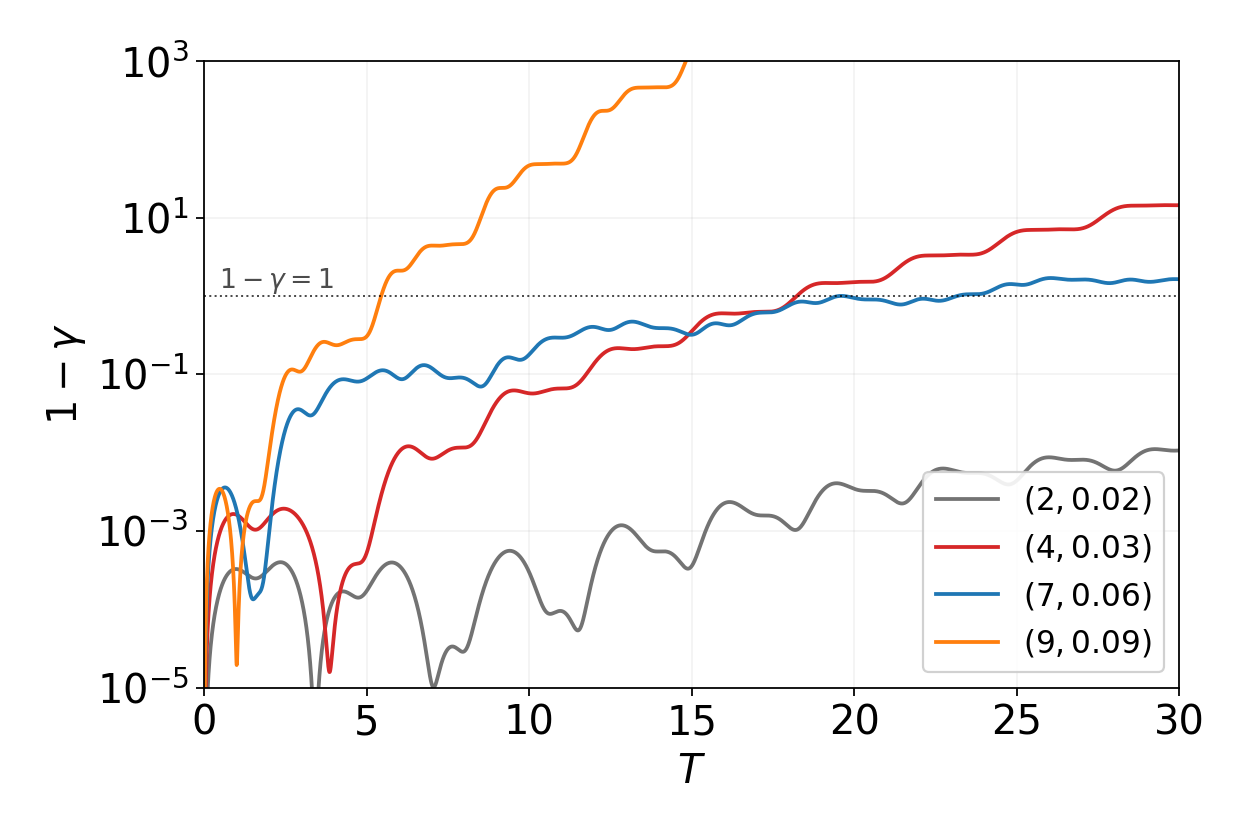}
    \caption{Plot of $1-\gamma$ on a logarithmic axis as a function of $T$ for $\omega=0.9$ at various points labeled by $(X_0,\lambda)$. The point $(4,0.03)$ is in the first resonance band and $(9,0.09)$ is in the second band. The threshold $1-\gamma=1$ is shown as the gray dotted line.}
    \label{fig:purityevol}
\end{figure}

\begin{figure}[h]
    \centering
    \includegraphics[width=0.9\linewidth]{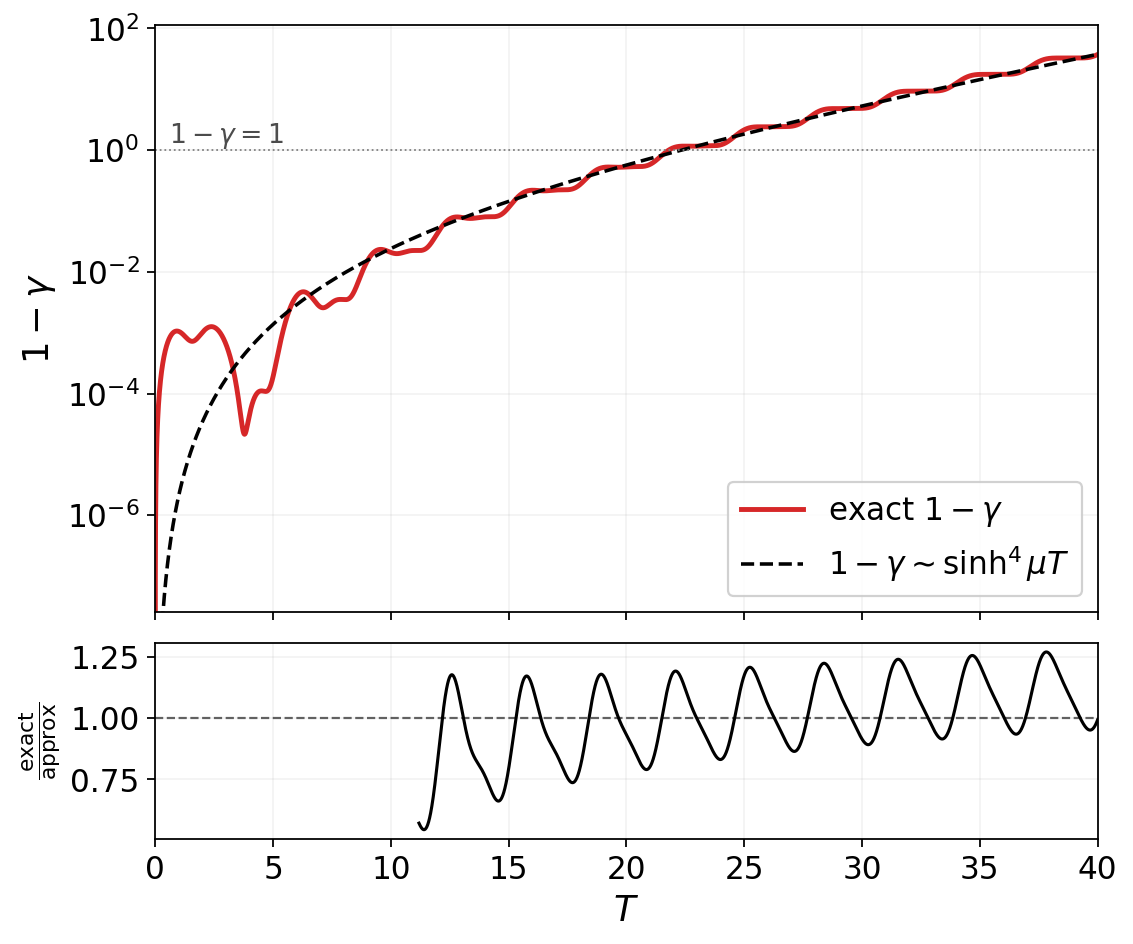}
    \caption{Comparison between the exact evolution of the purity for $\omega=0.9$ with the approximate late-time analytic form (\ref{eq:puritylatetime}). The threshold $1-\gamma=1$ is shown as the gray dotted line. The bottom plot gives the late-time ratio of these two quantities.}
    \label{fig:puritybandlaw}
\end{figure}

The time evolution of the purity is more explicitly shown in Fig.~\ref{fig:purityevol} for four points throughout the parameter space (for $\omega=0.9$). Two of these points, shown by the gray and blue lines, are outside the resonance bands and exhibit non-exponential growth. For the gray curve, corresponding to a point in the parametrically stable lower left corner of the $(X_0,\lambda )$ plane, the purity seems to follow a decreasing power law. In the case of the blue curve, corresponding to a point between the first and second resonance bands, the purity decreases sharply at the very beginning, closely mimicking the shape of the curve for the point in the second band (orange line) before settling on a more tame, power-law-like behavior. This could be due to it being closer to the ``strong-collective-coupling'' regime since, indeed, $N\tilde{\lambda}\approx 3$ in this case. For the orange and red curves we observe the exponential decrease of the purity associated with the instability bands. The behavior of the purity in the first band is more closely examined in Fig. ~\ref{fig:puritybandlaw}, where we compare the computed numerical evolution from Eq.~(\ref{eq:puritysimpl}) to the approximate form given by Eq.~(\ref{eq:puritylatetime}). Remarkably, the approximate analytical formula seems to agree quite well with the exact result as the curves closely follow each other (except at very early times). 

\begin{figure*}[ht]
    \centering
    \subfigure[]{\includegraphics[width=0.45\linewidth]{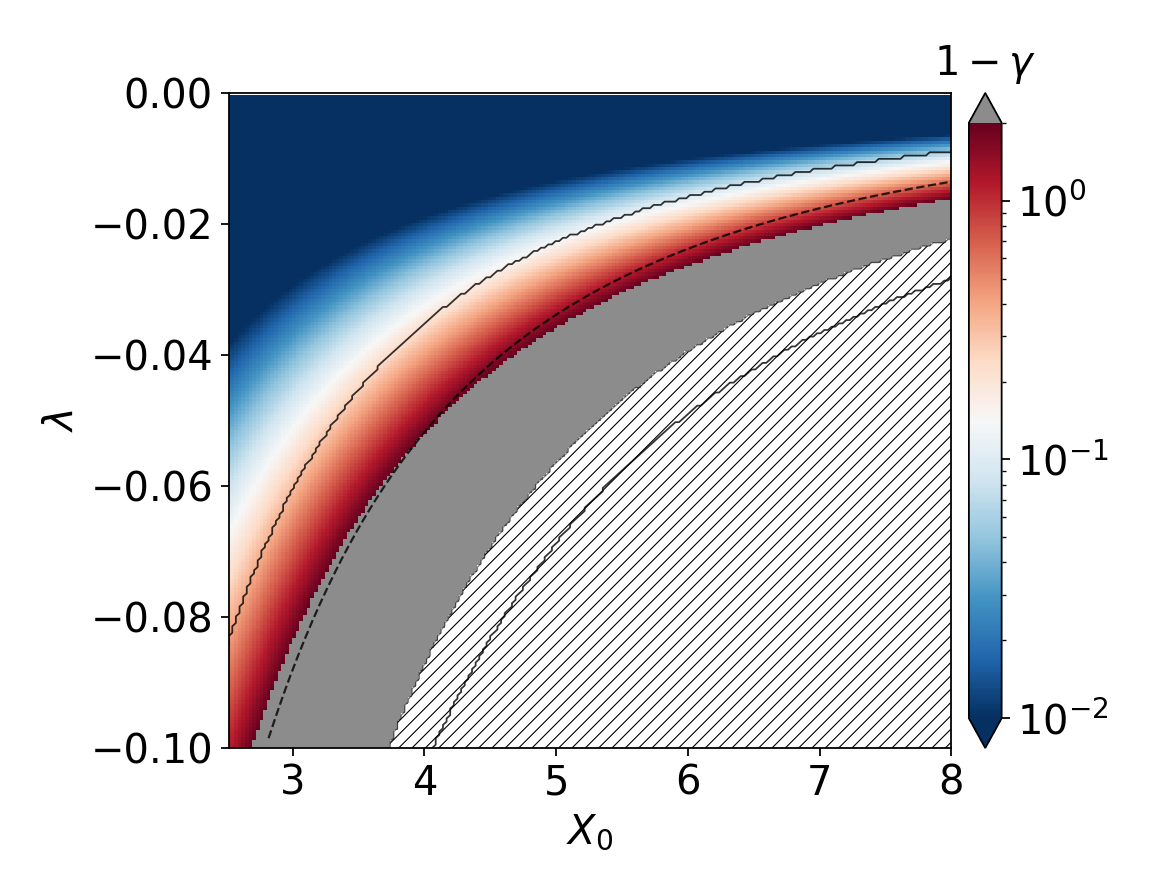}}\qquad
    \subfigure[]{\includegraphics[width=0.45\linewidth]{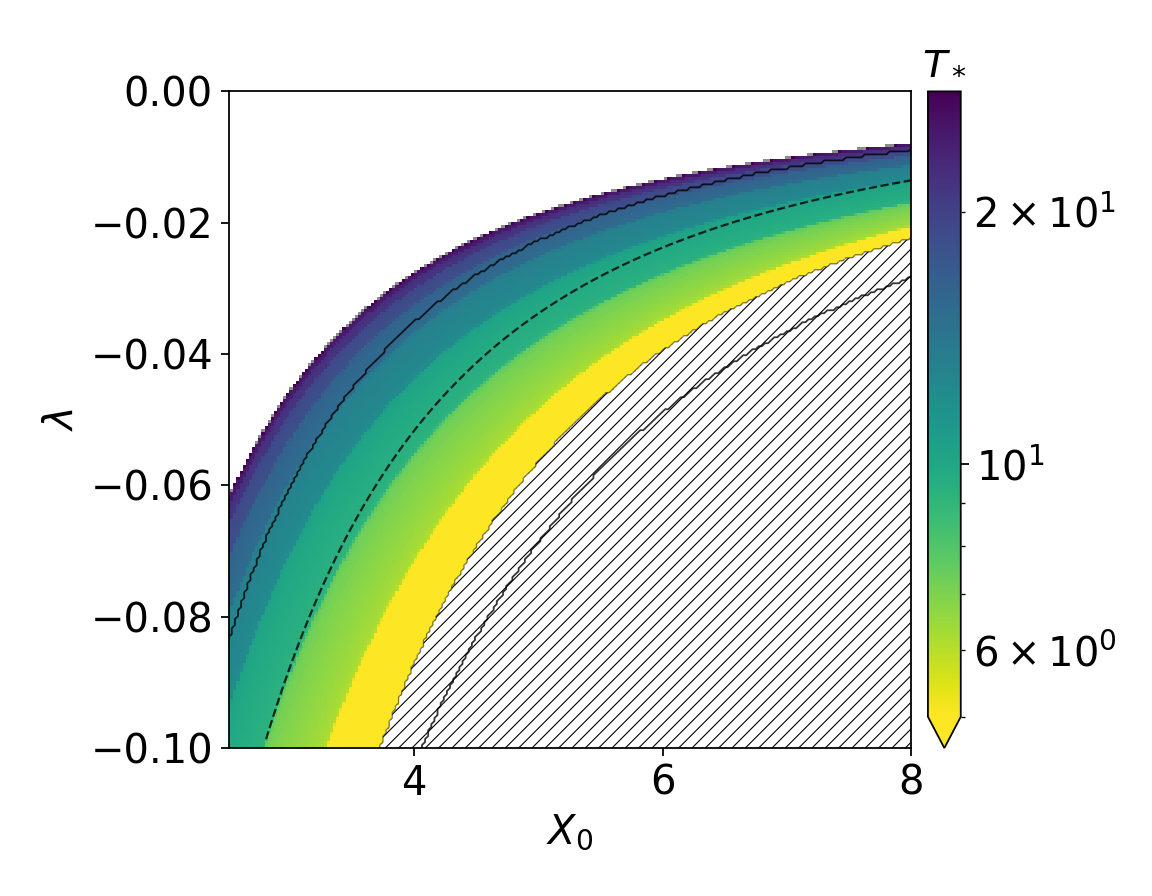}}
    \caption{(a) Density plot of $1-\gamma$ as a function of $X_0$ and $\lambda$ for  $\omega=1.2$ and $T=10$. The center of the resonance bands is shown with a dotted line and the edges with a solid line. Gray regions correspond to values of $1-\gamma\gtrsim1$. The lined region corresponds to the unphysical regime where $\omega_r^2<0$. (b) Density plot of the cutoff time $T_*$ for $\omega=1.2$. The white region means the cutoff value $1-\gamma=1$ is not reached within $T=30$. }
    \label{fig:omega1.2purity}
\end{figure*}

Finally, Fig. \ref{fig:omega1.2purity} also shows the behavior of $1-\gamma$, now for $\omega=1.2$. When $\omega>\Omega$, as is the case here, the first resonance band is defined by negative values of $\lambda$. We again see the expected peak in the resonance band, but there is now an additional feature. The lined region of the plot corresponds to a tachyonic instability, where $\omega_r^2<0$. In this regime, the purity would become complex.

\section{Conclusion}
\label{sec:conclusion}

We computed the purity for a system of coupled oscillators undergoing parametric resonance as a simplified model representing a classical background field coupled to a quantum matter field. The effect of the ``classical'' system  on the ``quantum'' system was taken into account using the Feynman-Vernon influence functional formalism. We used the purity as a measure of entanglement within the system and found, as expected, that purity decreases---and hence, entanglement increases---significantly when the system exhibits parametric resonance. This behavior is clearly seen in the first resonance band for various values of the frequency $\omega$. In fact, the growth of the quantity $1-\gamma$ (sometimes called {\it linear entropy}) was seen in Eq.~\eqref{eq:puritylatetime} to be exponential in time with a rate given by the Lyapunov exponent of the system. This is indeed not contradictory with the results of~\cite{Bianchi_2018,Zurek:1994wd} since they pertain to the entanglement entropy, which is expected to be  logarithmically suppressed compared to the linear entropy.

Our results show the background, approximately-classical, degree of freedom becoming more quantum over time due to quantum backreaction. This behavior was noted in \cite{Vachaspati_2017} for a similar system involving many degrees of freedom. It is usually thought that interactions tend to cause the quantum degree of freedom to become more classical. Here, however, we observe the opposite behavior since we explicitly probe a regime where quantum effects strongly entangle the background to the quantum degree of freedom and there is no mechanism to ensure decoherence.

Our results support previous research showing that the semiclassical approximation rapidly breaks down in the presence of parametric resonance \cite{van_Dissel_2025}, and we can surmise that it is likely due to the rapid increase in entanglement. This has ramifications for early universe cosmology, where the semiclassical approximation is often used. Indeed, as we have shown, our semiclassical model is only reliable up to $T\sim10$ (depending on the frequency used) before we get runaway secular behavior. This may have important consequences on predictions made for systems with parametric resonance.

It would be interesting to push the computations done in this work to the next order in perturbation theory and thus probe the robustness of our results. Another avenue would be to try to compute a non-perturbative quantity such as the entanglement entropy, but this would certainly involve a different set of methods. We leave such questions for further investigation.

\section{Acknowledgements}

The authors would like to thank Fabio van Dissel, Tanmay Vachaspati, Omkar Shetye, Chunshan Lin, and Moayad Al Khaled for useful discussions. RT is supported by the grant No. 2023/50/E/ST2/00231 from the National Science Centre, Poland. GZ is supported by SONATA BIS grant 2023/50/E/ST2/00231 from the National Science Centre, Poland.

\bibliography{refs}
\bibliographystyle{unsrt}

\end{document}